\documentclass{article}

\usepackage[preprint]{neurips_2026}

\usepackage[utf8]{inputenc} 
\usepackage[T1]{fontenc}    
\usepackage{hyperref}       
\usepackage{url}            
\usepackage{booktabs}       
\usepackage{amsfonts}       
\usepackage{nicefrac}       
\usepackage{microtype}      
\usepackage{xcolor}         

\usepackage{verbatim}
\usepackage{graphicx}
\usepackage{amssymb}
\usepackage{amsmath}
\usepackage{multirow}
\usepackage{algorithm2e}
\SetKwComment{Comment}{/* }{ */}
\RestyleAlgo{ruled}

\newcommand{\ie}{\textit{i.e.}}
\newcommand{\eg}{\textit{e.g.}}

\newcommand{\red}[1]{\textcolor{red}{#1}}
\newcommand{\blue}[1]{\textcolor{blue}{#1}}

\usepackage{pifont}
\newcommand{\cmark}{\ding{51}}
\newcommand{\xmark}{\ding{55}}

\title{GRAFT: Biological Graph and Hypergraph Benchmarks for Linked Gene Expression and Phenotypic Trait Prediction in \textit{Arabidopsis thaliana}}

\author{
  \textbf{Manuel Serna-Aguilera}$^{1, 6}$, 
  \textbf{Vanshika Jindal}$^{2}$, 
  \textbf{Fiona L. Goggin}$^{2}$, 
  \textbf{Jiamei Li}$^{2}$, \\
  \textbf{Aranyak Goswami}$^{3}$, 
  \textbf{Alexander Bucksch}$^{4}$, 
  \textbf{Suxing Liu}$^{5}$, 
  \textbf{Khoa Luu}$^{1, 6}$, \\
  $^{1}$Department of Electrical Engineering and Computer Science, University of Arkansas, AR \\
  $^{2}$Department of Entomology and Plant Pathology, University of Arkansas, AR \\
  $^{3}$Department of Animal Science, University of Arkansas, AR \\
  $^{4}$School of Plant Sciences, University of Arizona, Tucson, AZ \\
  $^{5}$Georgia State University, GA \\
  $^{6}$CVIU Lab, University of Arkansas, AR \\
  \small{\texttt{\{mserna, vjindal, fgoggin, jxl080, garanyak, khoaluu\}@uark.edu}} \\
  \small{\texttt{bucksch@arizona.edu}} \quad
  \small{\texttt{sliu58@gsu.edu}}
}

\begin{document}

\maketitle

\begin{abstract}
Understanding which genes control which traits in an organism remains one of the central challenges in biology. Despite significant advances in data collection technology, our ability to map genes to traits is still limited. This genome-to-phenome (G2P) challenge spans several problem domains, including plant breeding, and requires methods capable of reasoning over high-dimensional, heterogeneous, and biologically structured data. Current datasets and data repositories, however, are not well-equipped for this task. Current studies do not link gene expression and trait data, and most focus on very specific traits, limiting the breadth of possible correlations. 
    %
    %
To address this gap, we present the novel \textbf{G}ene-Graph \textbf{R}egression for \textbf{A}rabidopsis \textbf{F}unctional \textbf{T}raits (\textbf{GRAFT}) dataset, a curated multi-modal dataset linking gene expression profiles with phenotypic trait measurements in \textit{Arabidopsis thaliana}, a model organism in plant biology. GRAFT supports tasks such as phenotype prediction and interpretable graph learning. In addition, we benchmark conventional regression and explanatory baselines, including a biologically-informed hypergraph baseline, to validate gene-trait associations. To the best of our knowledge, this is the first dataset to provide multimodal gene information and heterogeneous trait or phenotype data for the same \textit{Arabidopsis thaliana} specimens. With GRAFT\footnote{All benchmark resources will be made publicly available upon acceptance.}, we aim to foster research to accurately understand the relationship between genotypes and phenotypes using gene information, higher-order gene pairings, and trait data from multiple sources. 
\end{abstract}

\section{Introduction}

Of the thousands of genes in an individual's genome and the hundreds of traits it displays, from their height to their health, which genes control which traits? The answers to this question are essential to nearly all applied life sciences, from crop improvement to animal breeding and medical drug development. Unfortunately, our ability to provide answers remains quite limited due to data analytics constraints. Decoding the relationship between an organism’s genetic makeup, \ie, its genome, and its traits, \ie, its phenome--the total of its different phenotypes, requires identifying complex patterns that interlink multiple high-dimensional and heterogeneous datasets. Unfortunately, there is a lack of benchmarking datasets to facilitate the development of computational tools, particularly in plant science. This hinders plant breeders’ efforts to meet increasing global food demands and combat emerging pests, diseases, and droughts. 

\noindent
\textbf{Limitations of Prior Work.} 
Research to crack the ``genome-to-phenome'' (G2P) challenge increasingly relies on high-dimensional and heterogeneous data to capture as many different aspects as possible of an individual specimen’s genome and phenome. 
Plant phenomic data typically has fewer features but is highly heterogeneous, ranging from images of shape and size to manual observations of development to spectrophotometric measurements of processes like photosynthesis. Combining data from more than one ``omics'' approach (\ie, multi-omics) is more effective at linking genes to specific phenotypes (observable traits) than any single omics approach alone \cite{minervini-2016, minervini-dataset-2015, ward-moghadam-dataset-2018, ward2019deepleafsegmentationusing, unl-ppd, arapheno-dataset-site, arapheno-seren-2017, photosynq-dataset-site}. Despite this, different types of plant omics data are siloed in separate repositories, as shown in Table \ref{tab:dataset-comparisons}, and there is a lack of comprehensive datasets that combine genomic and phenomic profiles from the same individuals to enable correlational analyses. Due to this lack of benchmarking data, the machine learning community has not kept up with the challenges of analyzing multi-omics data. Although there is much discussion in the biological literature about AI models that could infer correlations between samples in omics data \cite{cavill-omics-2015, cembrowska-omics-2023, demidchik-phenomics-2020, yang-omics-review-2021, gao-omics-short-review-2023, depuydt-omics-paper-2023, yan-ml-omics-review-2023, zhang-multi-omics-2022, mohammed-ai-omics-2023, wenhui-ml-omics-2024}, no such models seem to exist. Therefore, there is a critical need for biologically informed models that can map gene expression to heterogeneous phenotypes while enabling investigations into their explainability.

\noindent
\textbf{Problem Motivation.} To address knowledge gaps and limitations in both fields, we present the \textbf{G}ene-Graph \textbf{R}egression for \textbf{A}rabidopsis \textbf{F}unctional \textbf{T}raits (\textbf{GRAFT}), a dataset containing genomics (gene measurements) and phenomics (traits) data \textit{linked to the same specimens}, specifically of the foundational or ``model'' species \textit{Arabidopsis thaliana}. To go beyond gene-to-trait regression, GRAFT maps each gene to thousands of biological functions. Compared to other datasets, as shown in Table \ref{tab:dataset-comparisons}, GRAFT provides linked measurements not only for gene expression but also for many other traits typically examined in entire studies. 
To build biologically informed baselines, we use biologically-informed graphs and hypergraphs to represent $n$-order relationships among genes. 
We first benchmark regression models, and together with our team of biology experts, we explore 
SHapley Additive exPlanations (SHAP) \cite{lundberg-2017-shap-paper}, graph explanations \cite{ying2019gnnexplainergeneratingexplanationsgraph, luo2020parameterizedexplainergraphneural, zhang2024regexplainergeneratingexplanationsgraph}, in collaboration with biology experts, to construct a pipeline that aims to assist biologists in narrowing down top genes out of hundreds or thousands. Such a framework is beneficial in cutting labor costs and helping plant breeding efforts to zero-in on a small subset of important genes.

\noindent
\textbf{Contributions of this Work.} We summarize the contributions of this work as follows. \textbf{(\textit{i})} We contribute the novel GRAFT dataset, a collection of \textit{linked} gene expression and trait data. \textbf{(\textit{ii})} GRAFT additionally provides gene and biological function annotations that connect all genes, which we can interpret as hypergraphs. \textbf{(\textit{iii})} We provide insight into the explainability of the baseline regression methods, both quantitatively with our proposed biological explanation recall (BER) metric, and expert discussion. \textbf{(\textit{iv})} We will publicly release both our data and explanatory framework to foster further work into tackling the G2P challenge.


\section{Background and Related Work} \label{sec:background-related-work}

\begin{table}[t]\centering
\caption{
    Comparison of publicly-available datasets or data repositories, which lack diverse, multi-omics components compared to our dataset---GRAFT. GRAFT includes gene-level and heterogeneous phenotype-level information, while other common datasets and repositories do not. A blue check \blue{\cmark} indicates the dataset contains the corresponding measurement type, while \red{\xmark} indicates otherwise. 
}\label{tab:dataset-comparisons}
\scriptsize 
\begin{tabular}{lcccccc}
\textbf{} &\textbf{} &\textbf{Image-derived} &\textbf{Manual Measures of} &\textbf{Photosynthetic} &\textbf{Gene} \\
\textbf{Dataset/Repository} &\textbf{Images} &\textbf{Phenotypes} &\textbf{Growth/Development} &\textbf{Measurements} &\textbf{Expression} \\\midrule
I-PPD \cite{minervini-2016, minervini-dataset-2015} &\blue{\cmark} &\blue{\cmark} &\red{\xmark} &\red{\xmark} &\red{\xmark} \\
SAD \cite{ward-moghadam-dataset-2018, ward2019deepleafsegmentationusing} &\blue{\cmark} &\blue{\cmark} &\red{\xmark} &\red{\xmark} &\red{\xmark} \\
UNL-PPD \cite{unl-ppd} &\blue{\cmark} &\blue{\cmark} &\red{\xmark} &\red{\xmark} &\red{\xmark} \\
araPheno \cite{arapheno-dataset-site, arapheno-seren-2017} &\red{\xmark} &\red{\xmark} &\blue{\cmark} &\red{\xmark} &\red{\xmark} \\
Photosynq \cite{photosynq-dataset-site} &\red{\xmark} &\red{\xmark} &\red{\xmark} &\blue{\cmark} &\red{\xmark} \\
NCBI-GEO-SRA \cite{geo-repository-site, sra-site, ncbi-repository-2002} &\red{\xmark} &\red{\xmark} &\red{\xmark} &\red{\xmark} &\blue{\cmark} \\
ENA \cite{sra-site} &\red{\xmark} &\red{\xmark} &\red{\xmark} &\red{\xmark} &\blue{\cmark} \\
TAIR \cite{tair-site, tair-paper-2001, tallon-rnaseq-benchmarking-2024} &\red{\xmark} &\red{\xmark} &\red{\xmark} &\red{\xmark} &\blue{\cmark} \\
\midrule
\textbf{GRAFT} (Ours) &\blue{\cmark} &\blue{\cmark} &\blue{\cmark} &\blue{\cmark} &\blue{\cmark} \\
\bottomrule
\end{tabular}
\end{table}

\begin{figure*}
    \centering
    \includegraphics[width=0.75\linewidth]{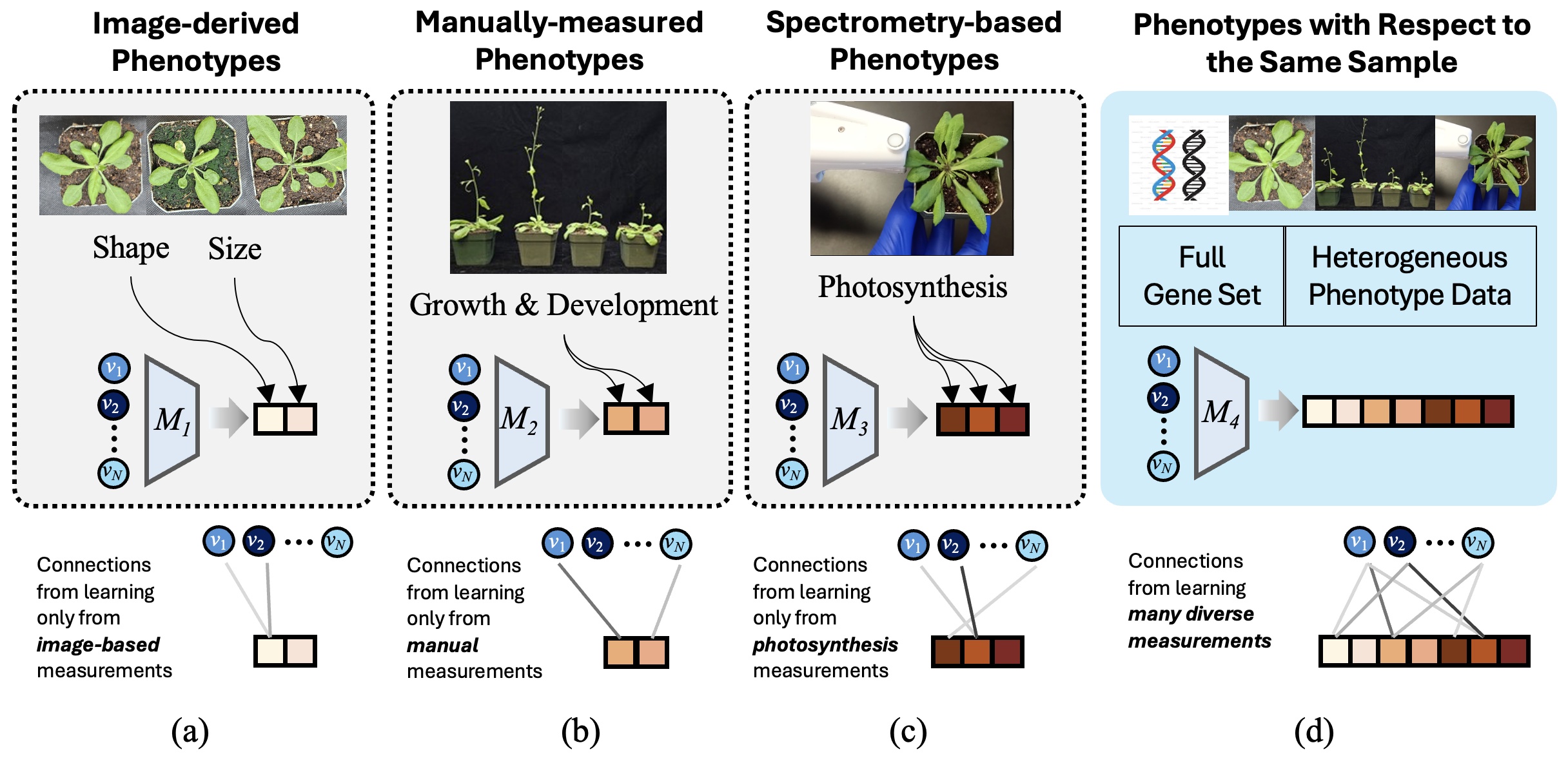}
    \caption{
        The problem that non-heterogeneous phenomics data poses to downstream tasks. 
        \textbf{(a)} Model $M_1$ is trained on homogeneous image-derived phenotypes on genes $\{v_1,\ldots,v_N\}$, and as a result cannot reason across other phenotypes. This is the inherent limitation of current datasets and benchmarks. 
        $M_2$ in case \textbf{(b)} suffers from similar limitations as $M_1$, so too does $M_3$ in \textbf{(c)}. 
        \textbf{(d)} With the breadth of phenotypes provided by GRAFT, $M_4$ learns to correlate genes to heterogeneous traits of the specimens.
        \textbf{Best viewed with zoom and in color.}
    }
    \label{fig:our-phenotype-data}
\end{figure*}

\subsection{The Model Plant \textit{Arabidopsis thaliana}}
\textit{Arabidopsis thaliana}, commonly known as the thale cress, is a model plant, much like rats and mice are ``model species'' that can help us learn about human health. The thale cress is widely used to study important crops such as corn and rice \cite{Woodward2018-hp}. The thale cress' relatively small genome and short lifecycle allowed for substantial research progress in the past several decades, making this species a natural choice for this work. In biology research, it is common to study mutant variants, often termed genotypic ``lines'' where at least one gene is mutated, leading to an alteration of some biological function, thereby motivating multi-omics studies \cite{cavill-omics-2015, cembrowska-omics-2023}. 


\subsection{Multi-Omics Analysis and the State of Omics Datasets}

``Omics'' or ``multi-omics'' data refers to measurements from different biological systems in an organism \cite{cavill-omics-2015, cembrowska-omics-2023}. 
Transcript\textit{omics} measures gene expression patterns, and RNA sequencing (RNA-seq) technology allows scientists to quantify almost all gene expressions \cite{mansoor-2025}. 
Gen\textit{omics}, is the study of the structure, function, and evolution of genomes within a given organism or community of organisms \cite{omics-definitions-paper-heavey-2022}. 
Phen\textit{omics} measures observable traits (examples in Section \ref{sec:dataset}). 
Omics data provides researchers with insight into which genes are activated during developmental stages or in response to specific environmental stimuli. Modern research is interested in multi-omics data because they provide a fuller picture of genes and traits, as shown in Fig. \ref{fig:our-phenotype-data}. 
Cembrowska et al. \cite{cembrowska-omics-2023} and many surveys \cite{demidchik-phenomics-2020, yang-omics-review-2021, gao-omics-short-review-2023, depuydt-omics-paper-2023, yan-ml-omics-review-2023, zhang-multi-omics-2022, mohammed-ai-omics-2023, wenhui-ml-omics-2024} provide further insight into omics research.

While high-throughput omics technology has improved considerably, most existing datasets fail to provide \textit{linked} multi-omics data, forcing studies to operate on limited data \cite{inferring-phenotypes-cheng-2021, flood-2016-circadian-rhythms, ubbens-deep-plant-phenomics-2017}. Major repositories (Table \ref{tab:dataset-comparisons}) often contain only phenotype or only expression data, but rarely both. 
Some datasets only provide image-derived data \cite{minervini-2016, minervini-dataset-2015, ward-moghadam-dataset-2018, ward2019deepleafsegmentationusing, unl-ppd}, or phenotype measurements \cite{arapheno-dataset-site, arapheno-seren-2017}, or photosynthetic data \cite{photosynq-dataset-site}, or gene-level data \cite{geo-repository-site, sra-site, ncbi-repository-2002, ebi-ena-site, tair-site, tair-paper-2001, tallon-rnaseq-benchmarking-2024}. None of these datasets links multi-omics data. Our GRAFT dataset, in contrast, addresses these gaps by offering complete gene expression profiles paired with rich phenotypic traits, enabling direct genotype-to-phenotype modeling. This positions our dataset as a valuable benchmark for developing scalable, integrative machine learning methods. Details on the datasets from Table \ref{tab:dataset-comparisons} are provided in the Appendix.

\subsection{Graph and Hypergraph Modeling}
Graph Convolutional Networks (GCNs) \cite{Bronstein_2017} take simple graphs, represented by node features and an edge list connecting pairs of nodes as its inputs, and produce embeddings of the interactions between nodes, \eg, via some message passing aggregation strategy, graph convolution \cite{kipf2017semisupervisedclassificationgraphconvolutional, defferrard2017convolutionalneuralnetworksgraphs, morris2021weisfeilerlemanneuralhigherorder}, graph attention \cite{velickovic2018graphattentionnetworks} or transformer operator \cite{shi2021maskedlabelpredictionunified}. Such networks also encode other kinds of data, such as images \cite{han2022visiongnnimageworth} or scene graphs \cite{nguyen2024highierarchicalinterlacementgraph, nguyen2024cyclocyclicgraphtransformer}. Recently, large language models have seen increasing use with knowledge graphs to provide grounded responses on large-scale databases \cite{choi2023nutreaneuraltreesearch, sun2024thinkongraphdeepresponsiblereasoning, chen2024llagalargelanguagegraph}, or even encoding subgraphs for such models \cite{perozzi2024letgraphtalkingencoding}. Graphs are also common in biology research for learning gene representations with graph structure \cite{liu2023musegnnlearningunifiedgene}, a similar but distinct topic to this work.

Hypergraph Neural Networks (HGNNs) utilize hypergraphs \cite{berge-hyper-textbook-1989, Bolla-1993, zien-1999, rodriguez2003, zhou-neurips2006, yadati-nerips2020} and encode $n$-ary relations, in which any set of $n$ nodes is connected by a hyperedge, rather than binary relations. In the past decade or so, deep learning has become integrated into hypergraph tasks, motivating the formulation of learnable hypergraph convolution \cite{yadati2019hypergcnnewmethodtraining, feng2019hypergraphneuralnetworks} or attention \cite{bai2020hypergraphconvolutionhypergraphattention}, and a variety of other investigations \cite{huang2021unignnunifiedframeworkgraph, alistarh-neurips2015, li2017inhomogeneoushypergraphclusteringapplications, kim2024hypeboygenerativeselfsupervisedrepresentation, wang2023hypergraphenergyfunctionshypergraph, wang2023equivarianthypergraphdiffusionneural}. For more practical tasks, hypergraphs have seen applications in scene generation \cite{nguyen2025hyperglmhypergraphvideoscene}, vision-natural language scenarios \cite{khan2023learningsituationhypergraphsvideo, kim-cvpr2020}, federated learning \cite{fan-cvpr2024}, recursive hyperedge structure \cite{yadati-nerips2020}, hypergraph matching \cite{zheng2024cursorscalablemixedorderhypergraph}, contrastive learning \cite{wei2022augmentationshypergraphcontrastivelearning}, recommendation systems \cite{han2022searchbehaviorpredictionhypergraph}, tabular data \cite{chen2023hytrelhypergraphenhancedtabulardata}, the long tail problem \cite{HAN2024112400, liu2023selfsuperviseddynamichypergraphrecommendation}, and integration with large language models \cite{feng2024graphslargelanguagemodels, chu2024llmguidedmultiviewhypergraphlearning, huang2025hyperghypergraphenhancedllmsstructured, luo2025hypergraphrag}. 
Some prior work \cite{hwang-hypergraph-genes-2008} has investigated hypergraphs to connect gene expression and protein interactions; this is distinct from our work, which focuses on gene-level and phenotype connections. Thus, this work provides a strong foundation for applying graphs and hypergraphs to the G2P challenge, as we will detail in Section \ref{sec:dataset}.

\subsection{Explanatory Methods and Explanatory Gene Discovery}
Explanatory methods seek to explain why a machine learning model makes particular predictions for a given input. They may explain the model prediction---why the model makes a certain prediction, or it may explain a ``phenomenon'' with respect to the nature of the data \cite{amara2024graphframexsystematicevaluationexplainability}. 
Well-known model-agnostic methods are LIME \cite{ribeiro2016whyitrustyou} and SHAP \cite{lundberg-2017-shap-paper}, which explain which inputs most contribute to the output. GNNExplainer \cite{ying2019gnnexplainergeneratingexplanationsgraph} PGExplainer \cite{luo2020parameterizedexplainergraphneural} offer efficient graph explanations, while RegExplainer \cite{zhang2024regexplainergeneratingexplanationsgraph} specifically addresses the graph regression task by incorporating a graph information bottleneck \cite{wu2020graphinformationbottleneck}. 

In the biology literature, several studies have examined correlations between genes and ``latent factor'' traits, but these studies are designed for different cases than ours. 
Approaches such as MOFA+ \cite{argelaguet_mofa_2020} and TotalVI \cite{gayoso_joint_2021} were designed for \textit{single-cell} multi-omics data integration, and focus on identifying correlations \textit{within individual cells} (\eg, single-cell RNAseq) and cell-level phenotypes \textit{within the same specimen}. Our work, in contrast, focuses on specimen-scale traits (\eg, plant morphology, photosynthesis, development) derived from bulk gene expression data and other whole-plant measurements. 
We focus on how gene expression patterns across an entire organism influence \textit{macroscopic} traits, rather than traits of a single cell. 
Similarly, DeepCCA \cite{pmlr-v28-andrew13}, while a powerful statistical tool for finding latent correlations between two datasets, is a general-purpose method that does not inherently incorporate the biologically-informed graph structures (\eg, incorporating gene-gene correlations into a graph) that are central to our framework's interpretability and its specific aim of understanding genotype-to-phenotype relationships within a biological context.

\section{The Proposed GRAFT Dataset} \label{sec:dataset}

We now discuss our proposed dataset---\textbf{G}ene-Graph \textbf{R}egression for \textbf{A}rabidopsis \textbf{F}unctional \textbf{T}raits (\textbf{GRAFT}). GRAFT contains gene data from multiple modalities, image-derived phenotype data, manually collected data, and spectrometry-based data. Section \ref{subsec:data-collection} discusses data collection, while Section \ref{subsec:mm-gene-anns} discusses our annotations for genes. Finally, Section \ref{subsec:go-and-hyperedge-encoding} discusses biological functions and their encoding into hyperedges.


\subsection{Data Collection} \label{subsec:data-collection}
This study utilizes four genetically distinct lines of \textit{Arabidopsis thaliana} (thale cress) that differ in gene expression and therefore in traits. These differences arise from targeted mutations in specific genes, making the lines well-suited for studying how genomic variation translates into observable phenotypic differences. We first collected image-based traits, manually-measured traits, and photosynthetic measurements using a spectrometer (Fig. \ref{fig:our-phenotype-data}). 
Following trait data collection, a subset of the same physical plants was destructively sampled for transcriptomic profiling via RNA sequencing (RNAseq), yielding \textbf{\textit{directly paired (or linked) genomic and phenotypic observations on the same individual specimens}}---a property that distinguishes this dataset from the overwhelming majority of published plant genomics resources (see Table \ref{tab:dataset-comparisons}). 
Further details on these genetic lines and important biological background are given in the Appendix. 

\subsubsection{Transcriptomics: Gene Expressions}
The GRAFT dataset provides whole-genome gene expression profiles convering $G=34,123$ transcripts currently annotated in the \textit{Arabidopsis thaliana} reference genome. For the purposes of this work, we can treat these are our genes.
This represents a near-complete transcriptomic coverage of a model plant genome in a benchmarking dataset. Gene expression values are derived from RNAseq analysis of 24 individual plant specimens (6 per line). We provide the gene measurements in Fragments Per Kilobase of transcript per Million mapped reads (FPKM) form. 


\subsubsection{Phenotypes: Observable Traits} \label{subsec:phenotype-level-info}
In total, phenotype data were collected for 77 individual thale cress plants drawn from four genetically distinct lines. Across these 77 plants, 41 phenotypic parameters were measured, spanning morphological, developmental, and physiological modalities. We focus on five parameters that collectively represent each measurement modality present in GRAFT and exhibit meaningful variation across the four lines. For the 24 plants that also underwent RNAseq profiling, all five parameters are directly paired with whole-genome expression measurements on the same individual, forming the core supervised learning pairs used for model training and evaluation. We do note one specimen must be dropped due to NaN values in its trait data, giving us 23 complete samples in experiments. Full details of the measurement protocols and per-line specimen counts are provided in the Appendix.

The five benchmark traits are: 
\textbf{(1) Leaf Area}: a manual measure of area covered by leaves in unit pixels (based on image analysis); 
\textbf{(2) Height} of the inflorescence or flower stalk: an indicator of how far into reproductive development a plant is (measured manually); 
\textbf{(3) FvP/FmP}: a measure of how efficiently the plant can channel light energy into photosynthesis (collected with a spectrometer); 
\textbf{(4) qL}: a measure of the chemical state of an important compound called plastoquinone in photosynthesis (also collected with a spectrometer); and 
\textbf{(5) Leaf temperature differential}: the difference in temperature between a leaf and its surroundings (also collected with a spectrometer). 



\subsection{Multimodal Gene Annotations} \label{subsec:mm-gene-anns}
\begin{figure*}
    \centering
    \includegraphics[width=0.95\linewidth]{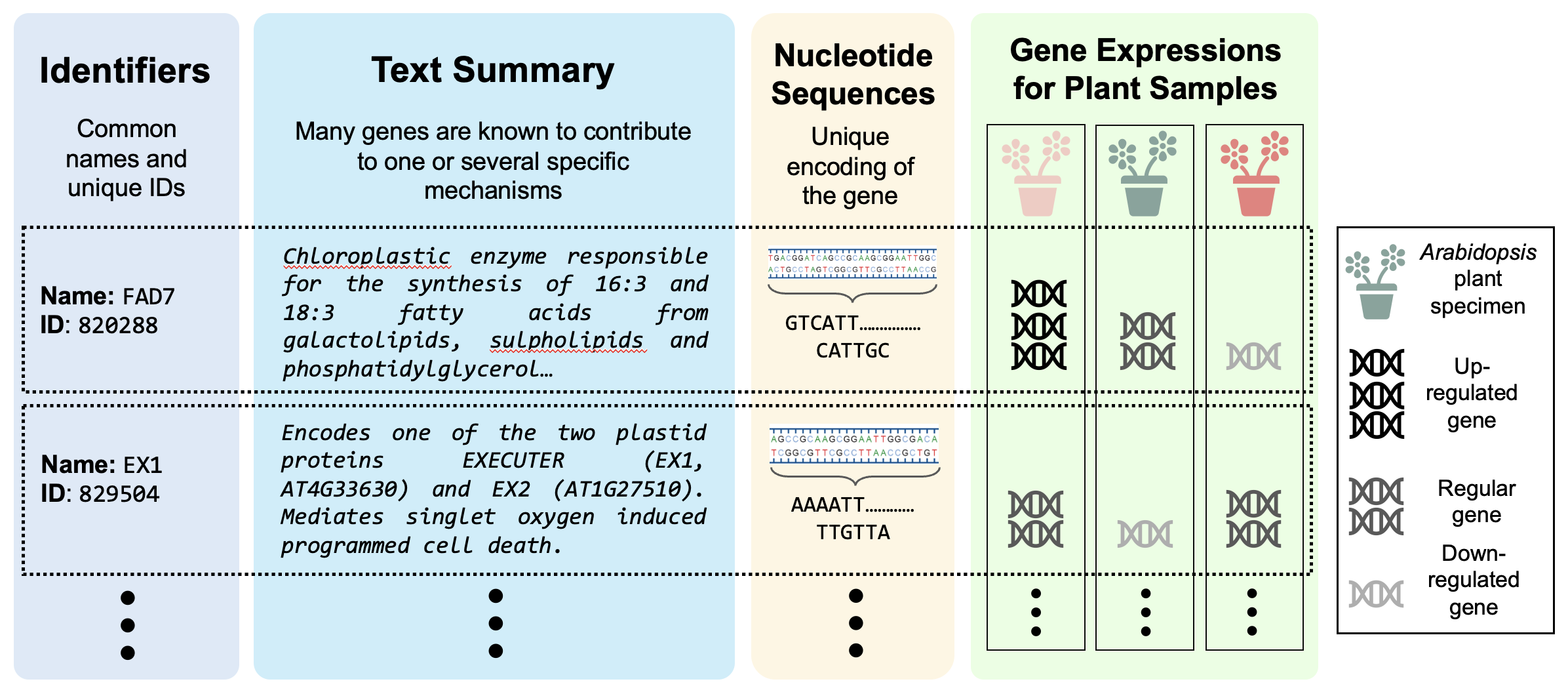}
    \caption{
        The gene-level features are provided by GRAFT. Genes have identifiers and text descriptions indicating what functions they influence. We also have the unique nucleotide sequence for each gene, the blueprint encoded as a string of characters. Finally, we add our gene expression data for all genes across different plant specimens. 
        \textbf{Best viewed with zoom and in color}.
    }
    \label{fig:gene-information}
\end{figure*}

Beyond raw expression values, GRAFT provides a rich annotation layer for every corresponding gene in the \textit{Arabidopsis thaliana} genome, enabling downstream interpretation of model outputs and explanations. Annotations were compiled from the National Center for Biotechnology Information (NCBI\footnote{\url{https://www.ncbi.nlm.nih.gov/datasets/gene/taxon/3702/}}) and include: (i) gene unique identifies (UIDs); (ii) common gene symbols (\eg, EX1, FAD7); and (iii) short descriptions. 
These annotations are organized in a tabular structure, allowing retrieval of functional descriptions for any gene or gene set of interest---for example, the set of high-importance genes identified by a model explanation method such as SHAP or a graph explanation. 
As illustrated in Fig. \ref{fig:gene-information}, we provide a diverse set of annotations per gene (not all used in this work).

This annotation layer is designed to close the interpretability gap that arises when machine learning models identify statistically important genes: rather than returning anonymous 
identifiers, users can immediately retrieve biological context for each gene, supporting faster hypothesis generation and experimental follow-up. 
For the GRAFT benchmark specifically, gene annotations support the Biological Explanation Recall (BER) evaluation metric (Section \ref{subsec:validation}), which measures whether a model's explanatory gene set recovers known biological functions relevant to predicted traits.

\begin{figure*}
    \centering
    \includegraphics[width=0.9\linewidth]{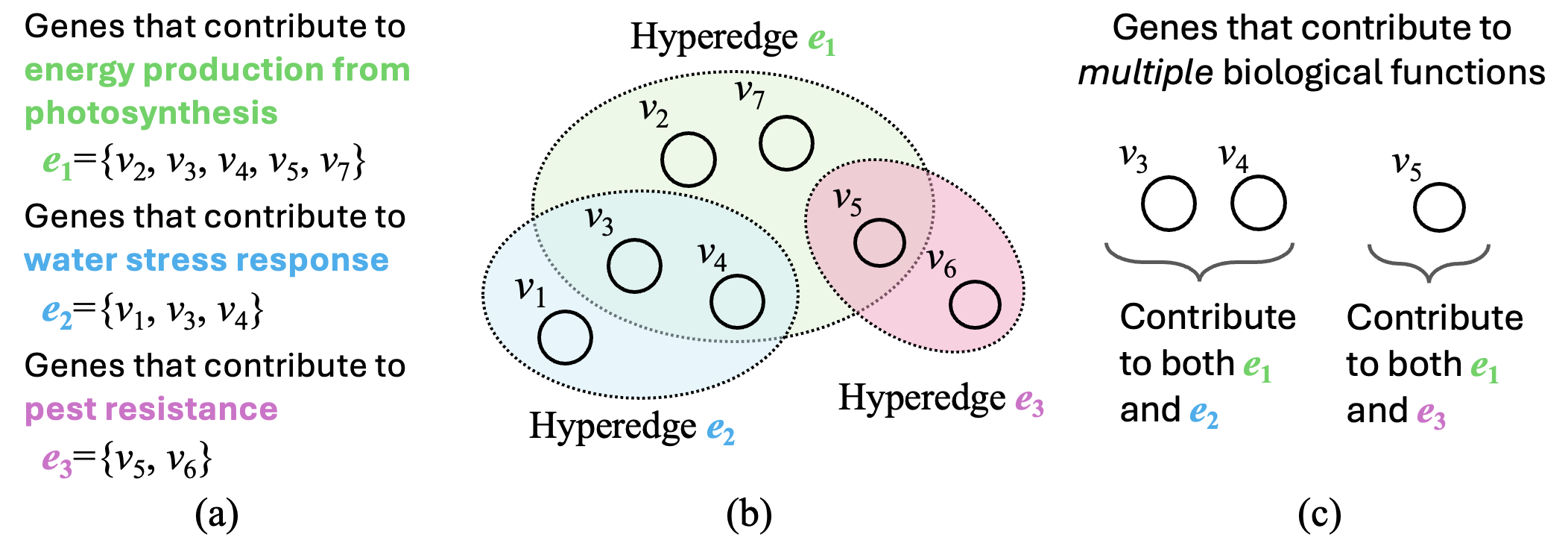}
    \caption{
        A visualization of translating the biological functions of the genes of the thale cress to higher-order pairings.
        \textbf{(a)} Thousands of functions exist for many systems in lifeforms, each containing genes that regulate those functions. 
        \textbf{(b)} The sets of genes can form hyperedges.
        \textbf{(c)} Hyperedges are connected by overlapping genes.
        \textbf{Best viewed with zoom and in color}.
    }
    \label{fig:hyperedge-relations}
\end{figure*}


\subsection{Gene Ontology Structure and Hyperedge Encoding} \label{subsec:go-and-hyperedge-encoding}
To capture the functional organization of the genome beyond individual gene annotations, GRAFT incorporates biological function information from the Gene Ontology (GO) resource\footnote{\url{https://geneontology.org/}}. ``GO'' terms provide descriptions of the molecular functions, biological processes, and cellular components that gene products are known or predicted to participate in. Each GO term carries a unique identifier, a standardized text description, and a set of gene associations indicating which genes are known to contribute to that function. 
For example, the GO term \texttt{GO:0047484} is described as \textit{``regulation of response to osmotic stress,''} and is associated with a known and established subset of genes.

The dataset includes annotations linking genes in our expression matrix to over 6,000 GO terms containing at least one gene. GO terms span a range of biological functions, both broad and fine-grained. Because each GO term defines a \textit{set} of genes that collectively participate in a shared function, GO terms map directly onto hyperedges in a hypergraph representation of the genome: each hyperedge connects the genes annotated to a given term, encoding biological co-functionality as graph structure as shown in Fig. \ref{fig:hyperedge-relations}. This hypergraph construction provides the structural prior used by our Hypergraph Neural Network (HGNN) baselines and enables biologically grounded explanations.

Taken together, these three annotation resources---per-gene identifiers and text annotations, GO term descriptions, and GO-to-gene hyperedges---form a self-contained biological knowledge base bundled with GRAFT. This design ensures that both the predictive and explanatory outputs of any model trained on GRAFT can be interpreted immediately in biological terms, without requiring users to query external databases separately.

\section{Methods} \label{sec:methodology}

\subsection{Preliminaries} \label{subsec:preliminaries}

Let us define a specimen $\mathcal{S}_i$'s linked data is defined as $(\mathbf{x}_i, \mathbf{y}_i)$, where $\mathbf{x}_i \in \mathbb{R}^{G\times1}$ holds all the $G\in\mathbb{Z}^+$ gene expressions for $\mathcal{S}_i$, and $\mathbf{y}_i \in \mathbb{R}^{T}$ denotes all $T\in\mathbb{Z}^+$ trait measurements of interest. There are $N$-many specimens. We learn a mapping $f\colon\mathbb{R}^{G}\to\mathbb{R}^{T}$
under a multi-output regression objective. Because $G \gg N$, we apply a two-stage \textit{gene filter} to prevent data leakage.

\paragraph{Gene Filtering.}
Within each cross-validation fold, using only training-set statistics:
(i) \textbf{\textit{Variance filter}}: genes with expression variance below $\tau_v = 0.01$ are discarded;
(ii) \textbf{\textit{Spearman filter}}: the top-$k$ genes ranked by maximum absolute Spearman correlation with any trait are retained, with $k=1024$. 
The filtered expression vector $\mathbf{z} \in \mathbb{R}^{k}$ is
the input to all downstream modeling. 
Further results with varying $k$ are provided in the Appendix.

\subsection{Graph-Based Regression} \label{subsec:graph-regression}

For graphs, we treat each gene as a node $v \in \mathcal{V}$ with initial feature $h_v^{(0)} = \mathbf{z}_v$ (its scalar expression value). Edges are given by the WGCNA \cite{pywgcna-2023} Topological Overlap Matrix (TOM), which encodes co-expression similarity derived from the same expression data used for prediction. Our key insight is that a biologically-derived adjacency, rather than a learned or arbitrary one, ensures that structural inductive biases align with \textit{known co-regulation patterns}, making subsequent explanations \textit{biologically interpretable}. In the graph case, we implement the gene-to-trait mapping as $\hat{\mathbf{y}} = f(\mathbf{z}; \mathbf{A})$, where $f$ consists of a GCN backbone and a prediction head, and $\mathbf{A}$ is the graph adjacency matrix. 

\subsection{Hypergraph-Based Regression} \label{subsec:hypergraph}

Pairwise edges cannot encode multi-gene biological pathways. To implement the HGNN, our key insight is to represent the genome as a hypergraph $\mathcal{H} = (\mathcal{V}, \mathcal{E}_H)$ where each hyperedge $e \in \mathcal{E}_H$ is the set of genes connected to a Gene Ontology (GO) term, encoded by incidence matrix $\mathbf{B} \in \{0,1\}^{k \times m}$, where $m$ indexes a GO term. This construction is a direct consequence of our dataset's GO annotation layer: every GO term becomes a hyperedge that groups genes by shared function or biological process, providing the HGNN with a pathway-level structural prior unavailable in purely pairwise graph formulations. In the hypergraph case, we implement the gene-to-trait mapping as $\hat{\mathbf{y}} = f(\mathbf{z}; \mathcal{H})$, where $f$ consists of an HGNN backbone and a prediction head.

\subsection{Explanations} \label{subsec:explanations}

A core goal of this benchmark is to assess not only predictive accuracy but also whether a model's rationale aligns with known biology. We obtain two complementary explanation signals.

\noindent
\textbf{SHAP.} For every trained $f$ we compute SHAP \cite{lundberg-2017-shap-paper} values (specifically using GradientSHAP, as DeepSHAP introduced complexities during implementation), yielding a per-gene, per-trait importance score $\phi_{g,t} \in \mathbb{R}$ that satisfies the Shapley efficiency axiom. Scores are averaged in absolute value across the test samples of each fold and then accumulated across folds in the original $G$-dimensional gene space, so that the final attribution $\bar{\phi}_{g,t}$ is comparable across models regardless of fold-varying feature subsets. The top-$k$ genes by $\bar{\phi}_{g,t}$ constitute the \textit{explanatory gene set} for trait $t$.

\noindent
\textbf{Graph explanations.}
To add another source of explanations, within the same analysis framework, for GCNs, we extract subgraph-level explanations via standardized graph explainers \cite{ying2019gnnexplainergeneratingexplanationsgraph, luo2020parameterizedexplainergraphneural, zhang2024regexplainergeneratingexplanationsgraph}. Discussion of this class of explanations is discussed in the Appendix.

\subsection{Explanation Validation} \label{subsec:validation}



\noindent
\textbf{Biological Explanation Recall (BER) Score.} For each trait $t$ and each model, the explanatory gene set is input to GO enrichment analysis (hypergeometric test \cite{klopfenstein_goatools_2018}). BER@$k$ is the fraction of \textit{a priori} trait-relevant GO terms recovered among the significantly enriched terms in Eqn \ref{eq:ber}, where $\mathcal{T}_{\mathrm{relevant}}$ is manually fixed before any model is run (exact GO terms are given in the Appendix).
\begin{equation}
  \mathrm{BER}@k \;=\;
  \frac{|\,\mathcal{T}_{\mathrm{enrich}} \cap \mathcal{T}_{\mathrm{relevant}}\,|}
       {|\mathcal{T}_{\mathrm{relevant}}|},
  \label{eq:ber}
\end{equation}
BER@$k$ scores are reported across $k \in \{50, 100, 200, 500\}$, $k$ denoting top-genes from SHAP, enabling comparison of explanation quality independently of predictive accuracy.

\noindent
\textbf{DEG List Overlap.} We compare each model type's top-$k$ SHAP genes against the differentially expressed gene (DEG) lists returned by statistical analysis of our expression data across the pairwise line comparisons. We note that this analysis does not relate to \textit{specific traits}, which motivates our own analysis. This validation is independent of GO annotations, thus, a gene identified by both analyses strengthens the gene identification confidence.

\section{Benchmarks and Evaluation} \label{sec:benchmarks-and-eval}

\subsection{Trait Regression}

\begin{table}\centering
\caption{Root mean-square error (rMSE) for stratified $K$-fold cross-validation results \textbf{\textit{for the top 1024 genes}} over three random seed splittings. Note: leaf temperature differential = \textbf{LTD}.
}\label{tab:kfold-regression}
\resizebox{\textwidth}{!}{ 
\begin{tabular}{lcccccc}\toprule
\multirow{2}{*}{\textbf{Backbone}} &\multicolumn{5}{c}{\textbf{rMSE ($\mu \pm \sigma^{2}$)}} \\\cmidrule{2-6}
&\textbf{Leaf Area} &\textbf{Inflorescence Height} &\textbf{qL} &\textbf{FvP/FmP} &\textbf{LTD} \\\cmidrule{1-6}
MLP                        & 1.058 $\pm$ 0.198 & 1.051 $\pm$ 0.136 & 1.002 $\pm$ 0.349 & 1.069 $\pm$ 0.256 & 0.942 $\pm$ 0.138 \\\cmidrule{1-6}
GCN + GraphConv \cite{morris2021weisfeilerlemanneuralhigherorder} & 1.197 $\pm$ 0.380 & 1.338 $\pm$ 0.589 & 1.336 $\pm$ 0.450 & 1.390 $\pm$ 0.555 & 1.128 $\pm$ 0.393 \\
GCN + TConv \cite{shi2021maskedlabelpredictionunified}     & 1.512 $\pm$ 0.367 & 1.684 $\pm$ 0.611 & 1.293 $\pm$ 0.250 & 1.824 $\pm$ 1.027 & 1.440 $\pm$ 0.629 \\
GCN + SAGEConv \cite{hamilton2018inductiverepresentationlearninglarge}  & 1.197 $\pm$ 0.380 & 1.338 $\pm$ 0.589 & 1.336 $\pm$ 0.450 & 1.390 $\pm$ 0.555 & 1.128 $\pm$ 0.393 \\\cmidrule{1-6}
HGNN + HConv \cite{bai2020hypergraphconvolutionhypergraphattention}    & 0.967 $\pm$ 0.208 & 1.014 $\pm$ 0.126 & 0.953 $\pm$ 0.295 & 0.960 $\pm$ 0.276 & 0.991 $\pm$ 0.111 \\
HGNN + HyperGCN \cite{yadati2019hypergcnnewmethodtraining} & 0.990 $\pm$ 0.243 & 1.003 $\pm$ 0.131 & 0.964 $\pm$ 0.324 & 0.983 $\pm$ 0.243 & 0.994 $\pm$ 0.099 \\
HGNN + UniGAT \cite{huang2021unignnunifiedframeworkgraph}   & 0.961 $\pm$ 0.217 & 0.999 $\pm$ 0.119 & 0.957 $\pm$ 0.322 & 0.963 $\pm$ 0.247 & 0.987 $\pm$ 0.118 \\
\bottomrule
\end{tabular}
}
\end{table}

We begin our benchmarking and evaluation on our dataset by testing several regression backbones with gene expressions as input, and report root mean square error (rMSE).
We begin with a baseline MLP with dense connections, then GCNs: GraphConv \cite{morris2021weisfeilerlemanneuralhigherorder}, graph transformer (TConv) \cite{shi2021maskedlabelpredictionunified}, and SAGE (SAGEConv) \cite{hamilton2018inductiverepresentationlearninglarge}; and HGNNs: Hypergraph convolution (HConv) \cite{bai2020hypergraphconvolutionhypergraphattention}, HyperGCN \cite{yadati2019hypergcnnewmethodtraining}, and UniGAT \cite{huang2021unignnunifiedframeworkgraph}.
To address the high-dimensionality and low-sample setting, we train the networks in three different cross-validation settings: leave-one-out (LOO), leave-one-class-out (LOCO), and stratified $K$-fold (SKFOLD). We report on SKFOLD in the manuscript, as in biological scenarios, we want models to train on all genetic lines or classes, LOO and LOCO results are not as biologically relevant. From Table \ref{tab:kfold-regression}, HGNNs consistently outperform MLPs and GCNs with their compact, biologically-informed, and interpretable hypergraph design. Exact training settings, as well as LOO and LOCO results are provided in the Appendix. 

\subsection{Explanation Validation} \label{subsec:explanation-validation}

\begin{table}[t!]
\small
\centering
\caption{
    Mean BER@$k$ across stratified folds and model types. (-) denotes 0.0$\pm$0.0.
}\label{tab:ber}
\begin{tabular}{llcccc}\toprule
\multirow{2}{*}{\textbf{Model}} & \multirow{2}{*}{\textbf{Trait}} & \multicolumn{4}{c}{BER@$k$} \\
\cmidrule(lr){3-6}
& & $k=50$ & $k=100$ & $k=200$ & $k=500$ \\
\midrule
MLP & Leaf Area & - & - & - & - \\
         & Inflorescence & - & - & - & 0.009$\pm$0.019 \\
         & qL & 0.026$\pm$0.043 & 0.026$\pm$0.043 & 0.013$\pm$0.026 & 0.052$\pm$0.108 \\
         & FvP/FmP & - & 0.016$\pm$0.031 & 0.032$\pm$0.063 & 0.056$\pm$0.112 \\
         & LTD & - & - & - & 0.007$\pm$0.020 \\
\midrule
GCN & Leaf Area & - & - & - & - \\
         & Inflorescence & - & - & 0.002$\pm$0.007 & 0.010$\pm$0.020 \\
         & qL & - & - & 0.004$\pm$0.023 & 0.052$\pm$0.118 \\
         & FvP/FmP & - & - & - & 0.066$\pm$0.120 \\
         & LTD & - & - & - & 0.004$\pm$0.016 \\
\midrule
HGNN & Leaf Area & 0.030$\pm$0.053 & 0.042$\pm$0.065 & 0.028$\pm$0.050 & 0.014$\pm$0.032 \\
         & Inflorescence & 0.043$\pm$0.046 & 0.059$\pm$0.067 & 0.065$\pm$0.066 & 0.038$\pm$0.042 \\
         & qL & 0.033$\pm$0.105 & 0.057$\pm$0.161 & 0.096$\pm$0.213 & 0.122$\pm$0.237 \\
         & FvP/FmP & 0.053$\pm$0.163 & 0.082$\pm$0.203 & 0.122$\pm$0.241 & 0.124$\pm$0.241 \\
         & LTD & 0.068$\pm$0.090 & 0.100$\pm$0.095 & 0.115$\pm$0.122 & 0.072$\pm$0.085 \\
\bottomrule
\end{tabular}
\end{table}

\begin{table}[t!]
\centering
\caption{
    DEG overlap for top-$k=1024$ genes.
}\label{tab:deg}
\begin{tabular}{lccccccc} 
\toprule
Trait & SAGEConv & TConv & GraphConv & HyperGCN & HConv & UniGAT & MLP \\
\midrule
Leaf area     & 1.110 & 0.610 & 1.110 & 0.150 & 10.306 & 2.209 & 0.263 \\
Inflorescence & 0.610 & 1.358 & 0.610 & 0.150 & 12.418 & 2.209 & 0.487 \\
FvP/FmP       & 0.610 & 0.610 & 0.610 & 0.150 & 28.436 & 1.159 & 0.487 \\
qL            & 0.487 & 1.358 & 0.487 & 0.110 & 5.374  & 2.209 & 0.276 \\
LTD           & 1.110 & 0.610 & 1.110 & 0.150 & 15.122 & 1.394 & 0.610 \\
\bottomrule
\end{tabular}
\end{table}

\textbf{Biological Explanation Recall (BER).} For each model type (MLP, GCN, HGNN), we report the average BER scores across the five traits. As shown in Table \ref{tab:ber}, HGNN achieves higher BER across all traits and k values, confirming that GO term hyperedges improve the biological coherence of model explanations. For photosynthetic traits (qL, FvP/FmP), BER increases monotonically with k and plateaus near $k=500$, consistent with large, densely annotated pathway gene sets; for morphological traits (Leaf Area, LTD), BER peaks at $k=100$ to $200$ and declines at $k=500$, indicating that explanatory signal is concentrated in a small coherent gene set and diluted by including more, ``irrelevant'', genes. 
GCN achieves near-zero BER at $k\leq200$ despite competitive regression accuracy, a pattern consistent with over-smoothing due to dense TOM adjacency. Pairwise co-expression edges encode topological proximity rather than functional membership, producing diffuse SHAP attributions that, biologically, do not align with GO pathways. 
The MLPs recover no signal for Leaf Area and LTD at any k. This reflects the polygenic nature of these traits. MLPs show a modest signal for the more concentrated photosynthetic traits. Together, these results demonstrate that predictive accuracy and explanation fidelity are dissociable, and that the structural prior, not the regression loss, determines whether a model's explanations are biologically interpretable.

\textbf{Overlap with Differentially Expressed Genes (DEGs).} This experiment uses the $-\log_{10}$ hypergeometric $p$-value of the overlap at our primary gene set with $k=1024$ for each backbone most relevant to each trait. 
Results are given in Table \ref{tab:deg}, where the statistical significance of overlap between each model's top-$k=1024$ explanatory genes and independently derived DEG lists.
The HConv backbone achieves strongly significant overlap across all traits ($p-\log_{10}$ ranges from 5.4 to 28.4), with the photosynthetic traits (Inflorescence, FvP/FmP) showing the highest values, indicating that GO-term hyperedge message passing concentrates explanatory signal on genes that classical differential expression analysis independently identifies as biologically important. 
The UniGAT backbone also achieves significant overlap across most traits, confirming that biologically informed models can drive improvement. HyperGCN, despite sharing the same input $(\mathbf{z}, \mathcal{H})$, exhibits near-zero overlap. The GCN and MLP produce low and largely non-significant overlap, reinforcing our prior conclusion with BER that pairwise co-expression edges and dense feature-only regression do not align model explanations with independently validated differential expression patterns.

\section{Conclusions} \label{sec:conclusions}

In this work, we introduced GRAFT---the first dataset, to our knowledge, that provides \textit{linked} whole-genome expression data and heterogeneous phenotypic trait measurements for the same model plant, \textit{Arabidopsis thaliana}, specimens. GRAFT also provides full gene and Gene Ontology annotations, enabling biologically informed graph and hypergraph modeling. 
Our benchmarking shows that biologically informed and structured models (HGNNs in particular) match or exceed dense MLP regression across diverse traits and cross-validation protocols, and, crucially, produce explanatory gene sets with higher biological fidelity, as measured by BER and by convergent overlap with independently derived DEG lists. 
These results demonstrate that incorporating prior biological knowledge directly into model architecture yields interpretability gains that purely data-driven approaches cannot replicate at this sample size, traits desired by the biology community. 
We hope GRAFT lowers the barrier for the machine learning and biology communities to engage with the G2P challenge, and that the benchmark tasks, evaluation metrics, and public code release provide a foundation for future work on explainable multi-omics regression in plant science and beyond.

\textbf{Limitations.} 
We should reiterate that, while the linked data spans 24 samples, the small sample size makes it considerably expensive and time-consuming to obtain linked data for a single specimen. Regardless, limits absolute regression performance and the statistical power of fold-level evaluations. Future work in machine learning and biology must account for these limitations. GRAFT is specific to \textit{Arabidopsis thaliana}, and while this species is the model plant in plant biology, the biology community's desire for transferability of trained models and explanatory gene sets to crops or other species remains an open question. Finally, BER scores are bounded by the completeness of GO annotations, which are uneven across the genome, and trait-GO term mappings made possible by TAIR are a reflection of current knowledge, which will change over time.


\bibliography{main}{}

@article{Bronstein_2017,
    title={Geometric Deep Learning: Going beyond Euclidean data},
    volume={34},
    ISSN={1558-0792},
    url={http://dx.doi.org/10.1109/MSP.2017.2693418},
    DOI={10.1109/msp.2017.2693418},
    number={4},
    journal={IEEE Signal Processing Magazine},
    publisher={Institute of Electrical and Electronics Engineers (IEEE)},
    author={Bronstein, Michael M. and Bruna, Joan and LeCun, Yann and Szlam, Arthur and Vandergheynst, Pierre},
    year={2017},
    month=jul, 
    pages={18–42}
}

@misc{defferrard2017convolutionalneuralnetworksgraphs,
    title={Convolutional Neural Networks on Graphs with Fast Localized Spectral Filtering}, 
    author={Michaël Defferrard and Xavier Bresson and Pierre Vandergheynst},
    year={2017},
    eprint={1606.09375},
    archivePrefix={arXiv},
    primaryClass={cs.LG},
    url={https://arxiv.org/abs/1606.09375}, 
}

@misc{kipf2017semisupervisedclassificationgraphconvolutional,
    title={Semi-Supervised Classification with Graph Convolutional Networks}, 
    author={Thomas N. Kipf and Max Welling},
    year={2017},
    eprint={1609.02907},
    archivePrefix={arXiv},
    primaryClass={cs.LG},
    url={https://arxiv.org/abs/1609.02907}, 
}

@misc{morris2021weisfeilerlemanneuralhigherorder,
      title={Weisfeiler and Leman Go Neural: Higher-order Graph Neural Networks}, 
      author={Christopher Morris and Martin Ritzert and Matthias Fey and William L. Hamilton and Jan Eric Lenssen and Gaurav Rattan and Martin Grohe},
      year={2021},
      eprint={1810.02244},
      archivePrefix={arXiv},
      primaryClass={cs.LG},
      url={https://arxiv.org/abs/1810.02244}, 
}

@misc{velickovic2018graphattentionnetworks,
      title={Graph Attention Networks}, 
      author={Petar Veličković and Guillem Cucurull and Arantxa Casanova and Adriana Romero and Pietro Liò and Yoshua Bengio},
      year={2018},
      eprint={1710.10903},
      archivePrefix={arXiv},
      primaryClass={stat.ML},
      url={https://arxiv.org/abs/1710.10903}, 
}

@misc{shi2021maskedlabelpredictionunified,
      title={Masked Label Prediction: Unified Message Passing Model for Semi-Supervised Classification}, 
      author={Yunsheng Shi and Zhengjie Huang and Shikun Feng and Hui Zhong and Wenjin Wang and Yu Sun},
      year={2021},
      eprint={2009.03509},
      archivePrefix={arXiv},
      primaryClass={cs.LG},
      url={https://arxiv.org/abs/2009.03509}, 
}

@misc{hamilton2018inductiverepresentationlearninglarge,
      title={Inductive Representation Learning on Large Graphs}, 
      author={William L. Hamilton and Rex Ying and Jure Leskovec},
      year={2018},
      eprint={1706.02216},
      archivePrefix={arXiv},
      primaryClass={cs.SI},
      url={https://arxiv.org/abs/1706.02216}, 
}

@misc{han2022visiongnnimageworth,
      title={Vision GNN: An Image is Worth Graph of Nodes}, 
      author={Kai Han and Yunhe Wang and Jianyuan Guo and Yehui Tang and Enhua Wu},
      year={2022},
      eprint={2206.00272},
      archivePrefix={arXiv},
      primaryClass={cs.CV},
      url={https://arxiv.org/abs/2206.00272}, 
}

@misc{liu2023musegnnlearningunifiedgene,
      title={MuSe-GNN: Learning Unified Gene Representation From Multimodal Biological Graph Data}, 
      author={Tianyu Liu and Yuge Wang and Rex Ying and Hongyu Zhao},
      year={2023},
      eprint={2310.02275},
      archivePrefix={arXiv},
      primaryClass={cs.LG},
      url={https://arxiv.org/abs/2310.02275}, 
}

@misc{choi2023nutreaneuraltreesearch,
      title={NuTrea: Neural Tree Search for Context-guided Multi-hop KGQA}, 
      author={Hyeong Kyu Choi and Seunghun Lee and Jaewon Chu and Hyunwoo J. Kim},
      year={2023},
      eprint={2310.15484},
      archivePrefix={arXiv},
      primaryClass={cs.CL},
      url={https://arxiv.org/abs/2310.15484}, 
}

@misc{sun2024thinkongraphdeepresponsiblereasoning,
      title={Think-on-Graph: Deep and Responsible Reasoning of Large Language Model on Knowledge Graph}, 
      author={Jiashuo Sun and Chengjin Xu and Lumingyuan Tang and Saizhuo Wang and Chen Lin and Yeyun Gong and Lionel M. Ni and Heung-Yeung Shum and Jian Guo},
      year={2024},
      eprint={2307.07697},
      archivePrefix={arXiv},
      primaryClass={cs.CL},
      url={https://arxiv.org/abs/2307.07697}, 
}

@misc{chen2024llagalargelanguagegraph,
      title={LLaGA: Large Language and Graph Assistant}, 
      author={Runjin Chen and Tong Zhao and Ajay Jaiswal and Neil Shah and Zhangyang Wang},
      year={2024},
      eprint={2402.08170},
      archivePrefix={arXiv},
      primaryClass={cs.LG},
      url={https://arxiv.org/abs/2402.08170}, 
}

@misc{perozzi2024letgraphtalkingencoding,
      title={Let Your Graph Do the Talking: Encoding Structured Data for LLMs}, 
      author={Bryan Perozzi and Bahare Fatemi and Dustin Zelle and Anton Tsitsulin and Mehran Kazemi and Rami Al-Rfou and Jonathan Halcrow},
      year={2024},
      eprint={2402.05862},
      archivePrefix={arXiv},
      primaryClass={cs.LG},
      url={https://arxiv.org/abs/2402.05862}, 
}

@inproceedings{lundberg-2017-shap-paper,
 author = {Lundberg, Scott M and Lee, Su-In},
 booktitle = {Advances in Neural Information Processing Systems},
 editor = {I. Guyon and U. Von Luxburg and S. Bengio and H. Wallach and R. Fergus and S. Vishwanathan and R. Garnett},
 pages = {},
 publisher = {Curran Associates, Inc.},
 title = {A Unified Approach to Interpreting Model Predictions},
 url = {https://proceedings.neurips.cc/paper_files/paper/2017/file/8a20a8621978632d76c43dfd28b67767-Paper.pdf},
 volume = {30},
 year = {2017}
}

@misc{ribeiro2016whyitrustyou,
      title={"Why Should I Trust You?": Explaining the Predictions of Any Classifier}, 
      author={Marco Tulio Ribeiro and Sameer Singh and Carlos Guestrin},
      year={2016},
      eprint={1602.04938},
      archivePrefix={arXiv},
      primaryClass={cs.LG},
      url={https://arxiv.org/abs/1602.04938}, 
}

@misc{ying2019gnnexplainergeneratingexplanationsgraph,
      title={GNNExplainer: Generating Explanations for Graph Neural Networks}, 
      author={Rex Ying and Dylan Bourgeois and Jiaxuan You and Marinka Zitnik and Jure Leskovec},
      year={2019},
      eprint={1903.03894},
      archivePrefix={arXiv},
      primaryClass={cs.LG},
      url={https://arxiv.org/abs/1903.03894}, 
}

@misc{luo2020parameterizedexplainergraphneural,
      title={Parameterized Explainer for Graph Neural Network}, 
      author={Dongsheng Luo and Wei Cheng and Dongkuan Xu and Wenchao Yu and Bo Zong and Haifeng Chen and Xiang Zhang},
      year={2020},
      eprint={2011.04573},
      archivePrefix={arXiv},
      primaryClass={cs.LG},
      url={https://arxiv.org/abs/2011.04573}, 
}

@misc{amara2024graphframexsystematicevaluationexplainability,
      title={GraphFramEx: Towards Systematic Evaluation of Explainability Methods for Graph Neural Networks}, 
      author={Kenza Amara and Rex Ying and Zitao Zhang and Zhihao Han and Yinan Shan and Ulrik Brandes and Sebastian Schemm and Ce Zhang},
      year={2024},
      eprint={2206.09677},
      archivePrefix={arXiv},
      primaryClass={cs.LG},
      url={https://arxiv.org/abs/2206.09677}, 
}

@misc{wu2020graphinformationbottleneck,
      title={Graph Information Bottleneck}, 
      author={Tailin Wu and Hongyu Ren and Pan Li and Jure Leskovec},
      year={2020},
      eprint={2010.12811},
      archivePrefix={arXiv},
      primaryClass={cs.LG},
      url={https://arxiv.org/abs/2010.12811}, 
}

@misc{zhang2024regexplainergeneratingexplanationsgraph,
      title={RegExplainer: Generating Explanations for Graph Neural Networks in Regression Tasks}, 
      author={Jiaxing Zhang and Zhuomin Chen and Hao Mei and Longchao Da and Dongsheng Luo and Hua Wei},
      year={2024},
      eprint={2307.07840},
      archivePrefix={arXiv},
      primaryClass={cs.LG},
      url={https://arxiv.org/abs/2307.07840}, 
}

@misc{feng2019hypergraphneuralnetworks,
      title={Hypergraph Neural Networks}, 
      author={Yifan Feng and Haoxuan You and Zizhao Zhang and Rongrong Ji and Yue Gao},
      year={2019},
      eprint={1809.09401},
      archivePrefix={arXiv},
      primaryClass={cs.LG},
      url={https://arxiv.org/abs/1809.09401}, 
}

@misc{bai2020hypergraphconvolutionhypergraphattention,
      title={Hypergraph Convolution and Hypergraph Attention}, 
      author={Song Bai and Feihu Zhang and Philip H. S. Torr},
      year={2020},
      eprint={1901.08150},
      archivePrefix={arXiv},
      primaryClass={cs.LG},
      url={https://arxiv.org/abs/1901.08150}, 
}

@misc{yadati2019hypergcnnewmethodtraining,
      title={HyperGCN: A New Method of Training Graph Convolutional Networks on Hypergraphs}, 
      author={Naganand Yadati and Madhav Nimishakavi and Prateek Yadav and Vikram Nitin and Anand Louis and Partha Talukdar},
      year={2019},
      eprint={1809.02589},
      archivePrefix={arXiv},
      primaryClass={cs.LG},
      url={https://arxiv.org/abs/1809.02589}, 
}

@INPROCEEDINGS{hwang-hypergraph-genes-2008,
  author={Hwang, TaeHyun and Tian, Ze and Kuangy, Rui and Kocher, Jean-Pierre},
  booktitle={2008 Eighth IEEE International Conference on Data Mining}, 
  title={Learning on Weighted Hypergraphs to Integrate Protein Interactions and Gene Expressions for Cancer Outcome Prediction}, 
  year={2008},
  volume={},
  number={},
  pages={293-302},
  doi={10.1109/ICDM.2008.37}
}

@book{berge-hyper-textbook-1989,
  title = {Graphs and hypergraphs},
  author = {Berge, Claude},
  year      = {1989},
  publisher = {Elsevier}
}

@ARTICLE{zien-1999,
  author={Zien, J.Y. and Schlag, M.D.F. and Chan, P.K.},
  journal={IEEE Transactions on Computer-Aided Design of Integrated Circuits and Systems}, 
  title={Multilevel spectral hypergraph partitioning with arbitrary vertex sizes}, 
  year={1999},
  volume={18},
  number={9},
  pages={1389-1399},
  doi={10.1109/43.784130}}

@misc{chen2023hytrelhypergraphenhancedtabulardata,
      title={HYTREL: Hypergraph-enhanced Tabular Data Representation Learning}, 
      author={Pei Chen and Soumajyoti Sarkar and Leonard Lausen and Balasubramaniam Srinivasan and Sheng Zha and Ruihong Huang and George Karypis},
      year={2023},
      eprint={2307.08623},
      archivePrefix={arXiv},
      primaryClass={cs.LG},
      url={https://arxiv.org/abs/2307.08623}, 
}

@inproceedings{yadati-nerips2020,
 author = {Yadati, Naganand},
 booktitle = {Advances in Neural Information Processing Systems},
 editor = {H. Larochelle and M. Ranzato and R. Hadsell and M.F. Balcan and H. Lin},
 pages = {3275--3289},
 publisher = {Curran Associates, Inc.},
 title = {Neural Message Passing for Multi-Relational Ordered and Recursive Hypergraphs},
 url = {https://proceedings.neurips.cc/paper_files/paper/2020/file/217eedd1ba8c592db97d0dbe54c7adfc-Paper.pdf},
 volume = {33},
 year = {2020}
}

@misc{zheng2024cursorscalablemixedorderhypergraph,
      title={CURSOR: Scalable Mixed-Order Hypergraph Matching with CUR Decomposition}, 
      author={Qixuan Zheng and Ming Zhang and Hong Yan},
      year={2024},
      eprint={2402.16594},
      archivePrefix={arXiv},
      primaryClass={cs.CV},
      url={https://arxiv.org/abs/2402.16594}, 
}

@INPROCEEDINGS{fan-cvpr2024,
  author={Fan, Q. and Shuai, L.},
  booktitle={2024 IEEE/CVF Conference on Computer Vision and Pattern Recognition (CVPR)}, 
  title={Adaptive Hyper-graph Aggregation for Modality-Agnostic Federated Learning}, 
  year={2024},
  volume={},
  number={},
  pages={12312-12321},
  doi={10.1109/CVPR52733.2024.01170}
}

@INPROCEEDINGS{kim-cvpr2020,
  author={Kim, Eun-Sol and Kang, Woo Young and On, Kyoung-Woon and Heo, Yu-Jung and Zhang, Byoung-Tak},
  booktitle={2020 IEEE/CVF Conference on Computer Vision and Pattern Recognition (CVPR)}, 
  title={Hypergraph Attention Networks for Multimodal Learning}, 
  year={2020},
  volume={},
  number={},
  pages={14569-14578},
  doi={10.1109/CVPR42600.2020.01459}
}

@misc{khan2023learningsituationhypergraphsvideo,
      title={Learning Situation Hyper-Graphs for Video Question Answering}, 
      author={Aisha Urooj Khan and Hilde Kuehne and Bo Wu and Kim Chheu and Walid Bousselham and Chuang Gan and Niels Lobo and Mubarak Shah},
      year={2023},
      eprint={2304.08682},
      archivePrefix={arXiv},
      primaryClass={cs.CV},
      url={https://arxiv.org/abs/2304.08682}, 
}

@inproceedings{zhou-neurips2006,
 author = {Zhou, Dengyong and Huang, Jiayuan and Sch\"{o}lkopf, Bernhard},
 booktitle = {Advances in Neural Information Processing Systems},
 editor = {B. Sch\"{o}lkopf and J. Platt and T. Hoffman},
 pages = {},
 publisher = {MIT Press},
 title = {Learning with Hypergraphs: Clustering, Classification, and Embedding},
 url = {https://proceedings.neurips.cc/paper_files/paper/2006/file/dff8e9c2ac33381546d96deea9922999-Paper.pdf},
 volume = {19},
 year = {2006}
}

@misc{wei2022augmentationshypergraphcontrastivelearning,
      title={Augmentations in Hypergraph Contrastive Learning: Fabricated and Generative}, 
      author={Tianxin Wei and Yuning You and Tianlong Chen and Yang Shen and Jingrui He and Zhangyang Wang},
      year={2022},
      eprint={2210.03801},
      archivePrefix={arXiv},
      primaryClass={cs.LG},
      url={https://arxiv.org/abs/2210.03801}, 
}

@article{rodriguez2003,
author = {J.A. Rodríguez},
title = {On the Laplacian Spectrum and Walk-regular Hypergraphs},
journal = {Linear and Multilinear Algebra},
volume = {51},
number = {3},
pages = {285--297},
year = {2003},
publisher = {Taylor \& Francis},
doi = {10.1080/0308108031000084374},
URL = {https://doi.org/10.1080/0308108031000084374},
eprint = {https://doi.org/10.1080/0308108031000084374}
}

@article{Bolla-1993,
	title = {Spectra, Euclidean representations and clusterings of hypergraphs},
	journal = {Discrete Mathematics},
	volume = {117},
	number = {1},
	pages = {19-39},
	year = {1993},
	issn = {0012-365X},
	doi = {https://doi.org/10.1016/0012-365X(93)90322-K},
	url = {https://www.sciencedirect.com/science/article/pii/0012365X9390322K},
	author = {Bolla, Marianna},
}

@inproceedings{alistarh-neurips2015,
 author = {Alistarh, Dan and Iglesias, Jennifer and Vojnovic, Milan},
 booktitle = {Advances in Neural Information Processing Systems},
 editor = {C. Cortes and N. Lawrence and D. Lee and M. Sugiyama and R. Garnett},
 pages = {},
 publisher = {Curran Associates, Inc.},
 title = {Streaming Min-max Hypergraph Partitioning},
 url = {https://proceedings.neurips.cc/paper_files/paper/2015/file/83f97f4825290be4cb794ec6a234595f-Paper.pdf},
 volume = {28},
 year = {2015}
}

@misc{li2017inhomogeneoushypergraphclusteringapplications,
      title={Inhomogeneous Hypergraph Clustering with Applications}, 
      author={Pan Li and Olgica Milenkovic},
      year={2017},
      eprint={1709.01249},
      archivePrefix={arXiv},
      primaryClass={cs.LG},
      url={https://arxiv.org/abs/1709.01249}, 
}

@misc{nguyen2025hyperglmhypergraphvideoscene,
      title={HyperGLM: HyperGraph for Video Scene Graph Generation and Anticipation}, 
      author={Trong-Thuan Nguyen and Pha Nguyen and Jackson Cothren and Alper Yilmaz and Khoa Luu},
      year={2025},
      booktitle={Conference on Computer Vision and Pattern Recognition},
      eprint={2411.18042},
      archivePrefix={arXiv},
      primaryClass={cs.CV},
      url={https://arxiv.org/abs/2411.18042}, 
}

@misc{nguyen2024cyclocyclicgraphtransformer,
      title={CYCLO: Cyclic Graph Transformer Approach to Multi-Object Relationship Modeling in Aerial Videos}, 
      author={Trong-Thuan Nguyen and Pha Nguyen and Xin Li and Jackson Cothren and Alper Yilmaz and Khoa Luu},
      year={2024},
      booktitle={Neural Information Processing Systems},
      eprint={2406.01029},
      archivePrefix={arXiv},
      primaryClass={cs.CV},
      url={https://arxiv.org/abs/2406.01029}, 
}

@misc{nguyen2024highierarchicalinterlacementgraph,
      title={HIG: Hierarchical Interlacement Graph Approach to Scene Graph Generation in Video Understanding}, 
      author={Trong-Thuan Nguyen and Pha Nguyen and Khoa Luu},
      year={2024},
      booktitle={Conference on Computer Vision and Pattern Recognition},
      eprint={2312.03050},
      archivePrefix={arXiv},
      primaryClass={cs.CV},
      url={https://arxiv.org/abs/2312.03050}, 
}

@misc{wang2023equivarianthypergraphdiffusionneural,
      title={Equivariant Hypergraph Diffusion Neural Operators}, 
      author={Peihao Wang and Shenghao Yang and Yunyu Liu and Zhangyang Wang and Pan Li},
      year={2023},
      eprint={2207.06680},
      archivePrefix={arXiv},
      primaryClass={cs.LG},
      url={https://arxiv.org/abs/2207.06680}, 
}

@misc{wang2023hypergraphenergyfunctionshypergraph,
      title={From Hypergraph Energy Functions to Hypergraph Neural Networks}, 
      author={Yuxin Wang and Quan Gan and Xipeng Qiu and Xuanjing Huang and David Wipf},
      year={2023},
      eprint={2306.09623},
      archivePrefix={arXiv},
      primaryClass={cs.LG},
      url={https://arxiv.org/abs/2306.09623}, 
}

@misc{kim2024hypeboygenerativeselfsupervisedrepresentation,
      title={HypeBoy: Generative Self-Supervised Representation Learning on Hypergraphs}, 
      author={Sunwoo Kim and Shinhwan Kang and Fanchen Bu and Soo Yong Lee and Jaemin Yoo and Kijung Shin},
      year={2024},
      eprint={2404.00638},
      archivePrefix={arXiv},
      primaryClass={cs.LG},
      url={https://arxiv.org/abs/2404.00638}, 
}

@misc{huang2021unignnunifiedframeworkgraph,
      title={UniGNN: a Unified Framework for Graph and Hypergraph Neural Networks}, 
      author={Jing Huang and Jie Yang},
      year={2021},
      eprint={2105.00956},
      archivePrefix={arXiv},
      primaryClass={cs.LG},
      url={https://arxiv.org/abs/2105.00956}, 
}

@misc{feng2024graphslargelanguagemodels,
      title={Beyond Graphs: Can Large Language Models Comprehend Hypergraphs?}, 
      author={Yifan Feng and Chengwu Yang and Xingliang Hou and Shaoyi Du and Shihui Ying and Zongze Wu and Yue Gao},
      year={2024},
      eprint={2410.10083},
      archivePrefix={arXiv},
      primaryClass={cs.AI},
      url={https://arxiv.org/abs/2410.10083}, 
}

@misc{chu2024llmguidedmultiviewhypergraphlearning,
      title={LLM-Guided Multi-View Hypergraph Learning for Human-Centric Explainable Recommendation}, 
      author={Zhixuan Chu and Yan Wang and Qing Cui and Longfei Li and Wenqing Chen and Zhan Qin and Kui Ren},
      year={2024},
      eprint={2401.08217},
      archivePrefix={arXiv},
      primaryClass={cs.IR},
      url={https://arxiv.org/abs/2401.08217}, 
}

@misc{huang2025hyperghypergraphenhancedllmsstructured,
      title={HyperG: Hypergraph-Enhanced LLMs for Structured Knowledge}, 
      author={Sirui Huang and Hanqian Li and Yanggan Gu and Xuming Hu and Qing Li and Guandong Xu},
      year={2025},
      eprint={2502.18125},
      archivePrefix={arXiv},
      primaryClass={cs.IR},
      url={https://arxiv.org/abs/2502.18125}, 
}

@misc{han2022searchbehaviorpredictionhypergraph,
      title={Search Behavior Prediction: A Hypergraph Perspective}, 
      author={Yan Han and Edward W Huang and Wenqing Zheng and Nikhil Rao and Zhangyang Wang and Karthik Subbian},
      year={2022},
      eprint={2211.13328},
      archivePrefix={arXiv},
      primaryClass={cs.IR},
      url={https://arxiv.org/abs/2211.13328}, 
}

@misc{liu2023selfsuperviseddynamichypergraphrecommendation,
      title={Self-Supervised Dynamic Hypergraph Recommendation based on Hyper-Relational Knowledge Graph}, 
      author={Yi Liu and Hongrui Xuan and Bohan Li and Meng Wang and Tong Chen and Hongzhi Yin},
      year={2023},
      eprint={2308.07752},
      archivePrefix={arXiv},
      primaryClass={cs.IR},
      url={https://arxiv.org/abs/2308.07752}, 
}

@article{HAN2024112400,
    title = {Dual-branch network with hypergraph feature augmentation and adaptive logits adjustment for long-tailed visual recognition},
    journal = {Applied Soft Computing},
    volume = {167},
    pages = {112400},
    year = {2024},
    issn = {1568-4946},
    doi = {https://doi.org/10.1016/j.asoc.2024.112400},
    url = {https://www.sciencedirect.com/science/article/pii/S1568494624011748},
    author = {Jia-yi Han and Jian-wei Liu and Jing-dong Xu}
}

@misc{luo2025hypergraphrag,
      title={HyperGraphRAG: Retrieval-Augmented Generation via Hypergraph-Structured Knowledge Representation}, 
      author={Haoran Luo and Haihong E and Guanting Chen and Yandan Zheng and Xiaobao Wu and Yikai Guo and Qika Lin and Yu Feng and Zemin Kuang and Meina Song and Yifan Zhu and Luu Anh Tuan},
      year={2025},
      eprint={2503.21322},
      archivePrefix={arXiv},
      primaryClass={cs.AI},
      url={https://arxiv.org/abs/2503.21322}, 
}

@ARTICLE{Woodward2018-hp,
  title     = "Biology in Bloom: A Primer on the Arabidopsis thaliana Model
               System",
  author    = "Woodward, Andrew W and Bartel, Bonnie",
  journal   = "Genetics",
  publisher = "Oxford University Press (OUP)",
  volume    =  208,
  number    =  4,
  pages     = "1337--1349",
  month     =  apr,
  year      =  2018,
  copyright = "https://academic.oup.com/journals/pages/open\_access/funder\_policies/chorus/standard\_publication\_model",
  language  = "en"
}

@article{flood-2016-circadian-rhythms,
    title = {Discovery and delivery strategies for engineered live biotherapeutic products},
    journal = {Plant Methods},
    volume = {12},
    number = {14},
    year = {2016},
    author = {Padraic J. Flood and Willem Kruijer and Sabine K. Schnabel and Rob van der Schoor and Henk Jalink and Jan F. H. Snel and Jeremy Harbinson and Mark G.M. Aarts},
}

@article{mansoor-2025,
	title = {Genomics, phenomics, and machine learning in transforming plant research: Advancements and challenges},
	journal = {Horticultural Plant Journal},
	volume = {11},
	number = {2},
	pages = {486-503},
	year = {2025},
	issn = {2468-0141},
	doi = {https://doi.org/10.1016/j.hpj.2023.09.005},
	url = {https://www.sciencedirect.com/science/article/pii/S2468014124000098},
	author = {Sheikh Mansoor and Ekanayaka M.B.M. Karunathilake and Thai Thanh Tuan and Yong Suk Chung}
}

@ARTICLE{ubbens-deep-plant-phenomics-2017,
	AUTHOR={Ubbens, Jordan R.  and Stavness, Ian },
	TITLE={Deep Plant Phenomics: A Deep Learning Platform for Complex Plant Phenotyping Tasks},
	JOURNAL={Frontiers in Plant Science},    
	VOLUME={Volume 8 - 2017},
	YEAR={2017},
	URL={https://www.frontiersin.org/journals/plant-science/articles/10.3389/fpls.2017.01190},
	DOI={10.3389/fpls.2017.01190},
	ISSN={1664-462X}
}

@article{inferring-phenotypes-cheng-2021,
    title = {Evolutionarily informed machine learning enhances the power of predictive gene-to-phenotype relationships},
    journal = {Nature Communications},
    volume = {12},
    number = {4567},
    year = {2021},
    url = {https://www.nature.com/articles/s41467-021-25893-w},
    author = {Chia-Yi Cheng and Ying Li and Kranthi Varala and Jessica Bubert and Ji Huang and Grace J. Kim and Justin Halim and Jennifer Arp and Hung-Jui S. Shih and Grace Levinson and Seo Hyun Park and Ha Young Cho and Stephen P. Moose and Gloria M. Coruzzi},
}

@article{omics-definitions-paper-heavey-2022,
    title = {Discovery and delivery strategies for engineered live biotherapeutic products},
    journal = {Trends in Biotechnology},
    volume = {40},
    number = {3},
    pages = {354-369},
    year = {2022},
    issn = {0167-7799},
    doi = {https://doi.org/10.1016/j.tibtech.2021.08.002},
    url = {https://www.sciencedirect.com/science/article/pii/S0167779921001761},
    author = {Mairead K. Heavey and Deniz Durmusoglu and Nathan Crook and Aaron C. Anselmo}
}

@article{cavill-omics-2015,
    author = {Cavill, Rachel and Jennen, Danyel and Kleinjans, Jos and Briedé, Jacob Jan},
    title = {Transcriptomic and metabolomic data integration},
    journal = {Briefings in Bioinformatics},
    volume = {17},
    number = {5},
    pages = {891-901},
    year = {2015},
    month = {10},
    issn = {1467-5463},
    doi = {10.1093/bib/bbv090},
    url = {https://doi.org/10.1093/bib/bbv090},
    eprint = {https://academic.oup.com/bib/article-pdf/17/5/891/6687226/bbv090.pdf},
}

@Article{cembrowska-omics-2023,
AUTHOR = {Cembrowska-Lech, Danuta and Krzemińska, Adrianna and Miller, Tymoteusz and Nowakowska, Anna and Adamski, Cezary and Radaczyńska, Martyna and Mikiciuk, Grzegorz and Mikiciuk, Małgorzata},
TITLE = {An Integrated Multi-Omics and Artificial Intelligence Framework for Advance Plant Phenotyping in Horticulture},
JOURNAL = {Biology},
VOLUME = {12},
YEAR = {2023},
NUMBER = {10},
ARTICLE-NUMBER = {1298},
URL = {https://www.mdpi.com/2079-7737/12/10/1298},
PubMedID = {37887008},
ISSN = {2079-7737},
DOI = {10.3390/biology12101298}
}

@article{demidchik-phenomics-2020,
    author = {V.V. Demidchik and
              A.Y. Shashko and
              U.Y. Bandarenka and
              G.N. Smolikova and
              D.A. Przhevalskaya and
              M.A. Charnysh and
              G.A. Pozhvanov and
              A.V. Barkosvkyi and
              I.I. Smolich and
              A.I. Sokolik and
              M. Yu and
              S.S. Medvedev},
    title = {Plant Phenomics: Fundamental Bases, Software and Hardware Platforms, and Machine Learning},
    journal = {Russian Journal of Plant Physiology },
    volume = {67},
    pages = {397-412},
    year = {2020},
}

@ARTICLE{yang-omics-review-2021,
    AUTHOR={Yang, Yaodong  and Saand, Mumtaz Ali  and Huang, Liyun  and Abdelaal, Walid Badawy  and Zhang, Jun  and Wu, Yi  and Li, Jing  and Sirohi, Muzafar Hussain  and Wang, Fuyou },
    TITLE={Applications of Multi-Omics Technologies for Crop Improvement},
    JOURNAL={Frontiers in Plant Science},
    VOLUME={12},
    YEAR={2021},
    URL={https://www.frontiersin.org/journals/plant-science/articles/10.3389/fpls.2021.563953},
    DOI={10.3389/fpls.2021.563953},
    ISSN={1664-462X},
}

@article{gao-omics-short-review-2023,
    author = {Gao, Feng and Huang, Kun and Xing, Yi},
    title = {Artificial Intelligence in Omics},
    journal = {Genomics, Proteomics \& Bioinformatics},
    volume = {20},
    number = {5},
    pages = {811-813},
    year = {2023},
    month = {01},
}

@article{depuydt-omics-paper-2023,
    author = {Depuydt, Thomas and De Rybel, Bert and Vandepoele, Klaas},
    title = {Charting plant gene functions in the multi-omics and single-cell era},
    journal = {Trends in Plant Science},
    volume = {28},
    issue = {3},
    pages = {283-296},
    year = {2023},
}

@article{yan-ml-omics-review-2023,
    author = {Yan, Jun and Wang, Xiangfeng},
    title = {Machine learning bridges omics sciences and plant breeding},
    journal = {Trends in Plant Science},
    volume = {28},
    issue = {2},
    pages = {199-210},
    year = {2023},
}

@article{zhang-multi-omics-2022,
    author = {Ru Zhang and Cuiping Zhang and Chengyu Yu and Jungang Dong and Jihong Hu},
    title = {Integration of multi-omics technologies for crop improvement: Status and prospects},
    journal = {Frontiers in bioinformatics},
    volume = {2},
    year = {2022},
}

@article{mohammed-ai-omics-2023, 
    title={A Comprehensive Review of Artificial Intelligence Approaches in Omics Data Processing: Evaluating Progress and Challenges}, 
    volume={2}, 
    url={https://ijmscs.org/index.php/ijmscs/article/view/8703}, 
    DOI={10.59543/ijmscs.v2i.8703}, 
    journal={International Journal of Mathematics, Statistics, and Computer Science}, 
    author={Ali, Ali Mahmoud and Mohammed, Mazin Abed}, 
    year={2023}, 
    month={Dec.}, 
    pages={114–167}
}

@article{wenhui-ml-omics-2024,
    author = {Bai, Wenhui and Li, Cheng and Li, Wei and Wang, Hai and Han, Xiaohong and Wang, Peipei and Wang, Li},
    title = {Machine learning assists prediction of genes responsible for plant specialized metabolite biosynthesis by integrating multi‑omics data},
    journal = {BMC Genomics},
    volume = {25},
    year = {2024},
}

@article{argelaguet_mofa_2020,
	title = {{MOFA}+: a statistical framework for comprehensive integration of multi-modal single-cell data},
	volume = {21},
	issn = {1474-760X},
	shorttitle = {{MOFA}+},
	url = {https://genomebiology.biomedcentral.com/articles/10.1186/s13059-020-02015-1},
	doi = {10.1186/s13059-020-02015-1},
	language = {en},
	number = {1},
	urldate = {2026-01-12},
	journal = {Genome Biology},
	author = {Argelaguet, Ricard and Arnol, Damien and Bredikhin, Danila and Deloro, Yonatan and Velten, Britta and Marioni, John C. and Stegle, Oliver},
	month = dec,
	year = {2020},
	pages = {111},
}

@article{gayoso_joint_2021,
	title = {Joint probabilistic modeling of single-cell multi-omic data with {totalVI}},
	volume = {18},
	issn = {1548-7091, 1548-7105},
	url = {https://www.nature.com/articles/s41592-020-01050-x},
	doi = {10.1038/s41592-020-01050-x},
	language = {en},
	number = {3},
	urldate = {2026-01-12},
	journal = {Nature Methods},
	author = {Gayoso, Adam and Steier, Zoë and Lopez, Romain and Regier, Jeffrey and Nazor, Kristopher L. and Streets, Aaron and Yosef, Nir},
	month = mar,
	year = {2021},
	pages = {272--282},
}

@InProceedings{pmlr-v28-andrew13,
  title = 	 {Deep Canonical Correlation Analysis},
  author = 	 {Andrew, Galen and Arora, Raman and Bilmes, Jeff and Livescu, Karen},
  booktitle = 	 {Proceedings of the 30th International Conference on Machine Learning},
  pages = 	 {1247--1255},
  year = 	 {2013},
  editor = 	 {Dasgupta, Sanjoy and McAllester, David},
  volume = 	 {28},
  number =       {3},
  series = 	 {Proceedings of Machine Learning Research},
  address = 	 {Atlanta, Georgia, USA},
  month = 	 {17--19 Jun},
  publisher =    {PMLR},
  url = 	 {https://proceedings.mlr.press/v28/andrew13.html}
}

@article{minervini-2016,
    title = {Finely-grained annotated datasets for image-based plant phenotyping},
    journal = {Pattern Recognition Letters},
    volume = {81},
    pages = {80-89},
    year = {2016},
    issn = {0167-8655},
    doi = {https://doi.org/10.1016/j.patrec.2015.10.013},
    url = {https://www.sciencedirect.com/science/article/pii/S0167865515003645},
    author = {Massimo Minervini and Andreas Fischbach and Hanno Scharr and Sotirios A. Tsaftaris}
}

@misc{minervini-dataset-2015,
    title = {Plant Phenotyping Datasets},
    howpublished = {\url{https://www.plant-phenotyping.org/datasets-home}},
    note = {Accessed: 2025-05-12}
}

@misc{ward-moghadam-dataset-2018,
    title = {Synthetic Arabidopsis Dataset},
    doi={https://doi.org/10.25919/5c36957c0af41},
    howpublished = {\url{https://doi.org/10.25919/5c36957c0af41}},
    note = {Accessed: 2025-05-12}
}

@misc{ward2019deepleafsegmentationusing,
      title={Deep Leaf Segmentation Using Synthetic Data}, 
      author={Daniel Ward and Peyman Moghadam and Nicolas Hudson},
      year={2019},
      eprint={1807.10931},
      archivePrefix={arXiv},
      primaryClass={cs.CV},
      url={https://arxiv.org/abs/1807.10931}, 
}

@misc{unl-ppd,
    title = {UNL Plant Phenotyping Datasets},
    howpublished = {\url{https://plantvision.unl.edu/unl-plant-phenotyping-datasets/}},
    note = {Accessed: 2025-05-12}
}

@misc{arapheno-dataset-site,
    title = {AraPheno},
    howpublished = {\url{https://arapheno.1001genomes.org/}},
    note = {Accessed: 2025-05-12}
}

@misc{arapheno-seren-2017,
      title={AraPheno: a public database for Arabidopsis thaliana phenotypes}, 
      Journal = {Nucleic Acids Research},
      author={Umit Seren and Dominik Grimm and Joffrey Fitz and Detlef Weigel and Magnus Nordborg and Karsten Borgwardt  and Arthur Korte},
      year={2017},
      doi={10.1093/nar/gkw986}
}

@misc{photosynq-dataset-site,
    title = {PhotosynQ},
    howpublished = {\url{https://photosynq.org/}},
    note = {Accessed: 2025-05-12}
}

@misc{geo-repository-site,
    title = {Gene Expression Omnibus},
    howpublished = {\url{https://www.ncbi.nlm.nih.gov/geo/}},
    note = {Accessed: 2025-05-12}
}

@misc{sra-site,
    title = {Sequence Read Archive},
    howpublished = {\url{https://www.ncbi.nlm.nih.gov/sra/}},
    note = {Accessed: 2025-05-12}
}

@misc{ebi-ena-site,
    title = {Sequence Read Archive},
    howpublished = {\url{https://www.ebi.ac.uk/ena/browser/home}},
    note = {Accessed: 2025-07-9}
}

@misc{ncbi-repository-2002,
      title={Gene Expression Omnibus: NCBI gene expression and hybridization array data repository}, 
      Journal = {Nucleic Acids Research},
      author={Ron Edgar and Michael Domrachev and Alex E Lash},
      year={2002},
      doi={10.1093/nar/30.1.207}
}

@misc{tair-site,
    title = {The Arabidopsis Information Resource},
    howpublished = {\url{https://www.arabidopsis.org/}},
    note = {Accessed: 2025-05-12}
}

@misc{tair-paper-2001,
      title={The Arabidopsis Information Resource (TAIR): a comprehensive database and web-based information retrieval, analysis, and visualization system for a model plant}, 
      Journal = {Nucleic Acids Research},
      author={E Huala and A W Dickerman and M Garcia-Hernandez and D Weems and L Reiser and F LaFond and D Hanley and D Kiphart and M Zhuang and W Huang and L A Mueller and D Bhattacharyya and D Bhaya and B W Sobral and W Beavis and D W Meinke and C D Town and C Somerville and S Y Rhee},
      year={2001},
      doi={https://doi.org/10.1093/nar/29.1.102}
}

@Article{tallon-rnaseq-benchmarking-2024,
    AUTHOR = {Coxe, Tallon and Burks, David J. and Singh, Utkarsh and Mittler, Ron and Azad, Rajeev K.},
    TITLE = {Benchmarking RNA-Seq Aligners at Base-Level and Junction Base-Level Resolution Using the Arabidopsis thaliana Genome},
    JOURNAL = {Plants},
    VOLUME = {13},
    YEAR = {2024},
    NUMBER = {5},
    ARTICLE-NUMBER = {582},
    URL = {https://www.mdpi.com/2223-7747/13/5/582},
    PubMedID = {38475429},
    ISSN = {2223-7747},
DOI = {10.3390/plants13050582}
}

@article{pywgcna-2023,
    author = {Rezaie, Narges and Reese, Farilie and Mortazavi, Ali},
    title = {PyWGCNA: a Python package for weighted gene co-expression network analysis},
    journal = {Bioinformatics},
    volume = {39},
    number = {7},
    pages = {btad415},
    year = {2023},
    month = {07},
    issn = {1367-4811},
    doi = {10.1093/bioinformatics/btad415},
    url = {https://doi.org/10.1093/bioinformatics/btad415},
    eprint = {https://academic.oup.com/bioinformatics/article-pdf/39/7/btad415/50920596/btad415.pdf},
}

@article{klopfenstein_goatools_2018,
	title = {{GOATOOLS}: {A} {Python} library for {Gene} {Ontology} analyses},
	volume = {8},
	issn = {2045-2322},
	shorttitle = {{GOATOOLS}},
	url = {https://www.nature.com/articles/s41598-018-28948-z},
	doi = {10.1038/s41598-018-28948-z},
	language = {en},
	number = {1},
	urldate = {2026-05-04},
	journal = {Scientific Reports},
	author = {Klopfenstein, D. V. and Zhang, Liangsheng and Pedersen, Brent S. and Ramírez, Fidel and Warwick Vesztrocy, Alex and Naldi, Aurélien and Mungall, Christopher J. and Yunes, Jeffrey M. and Botvinnik, Olga and Weigel, Mark and Dampier, Will and Dessimoz, Christophe and Flick, Patrick and Tang, Haibao},
	month = jul,
	year = {2018},
	pages = {10872},
}
\bibliographystyle{plain}

\end{document}